\documentclass[prl,showpacs,noshowkeys,twocolumn]{revtex4}
\usepackage{epsfig,amsfonts,amssymb,amsmath}
\usepackage{graphicx}
\usepackage{dcolumn}
\usepackage{bm}
\begin{document}
\title{Giant transversal particle diffusion in a longitudinal magnetic ratchet}
\author{Pietro Tierno$^1$}
\email{ptierno@ub.edu}
\author{Peter Reimann$^2$}
\author{Tom H. Johansen$^3$}
\author{Francesc Sagu\'{e}s$^1$}
\affiliation{
$^1$Departament de Qu\'{i}mica F\'{i}sica, Universitat de Barcelona, Mart\'{i} i Franqu$\grave{e}$s 1, 08028 Barcelona, Spain\\
$^2$Fakult\"{a}t f\"{u}r Physik, Universit\"{a}t Bielefeld, 33615 Bielefeld, Germany\\
$^3$Department of Physics, University of Oslo, P. O. Box 1048, Blindern, Norway}
\date{\today}
\begin{abstract}
We study the transversal motion of paramagnetic particles
on a uniaxial garnet film, exhibiting a longitudinal ratchet
effect in the presence of an oscillating magnetic field.
Without the field, the thermal diffusion coefficient obtained by video
microscopy is $D_{0}\approx 3\times 10^{-4}$ $\mu m^2/s$.
With the field, the transversal diffusion exhibits a giant enhancement by
almost four decades and a pronounced maximum as a function
of the driving frequency.
We explain the experimental findings with a theoretical interpretation 
in terms of random disorder effects
within the magnetic film.
\end{abstract}
\maketitle
Transport and diffusion
in periodic or random potentials play a key role
in many different contexts of Physics, Chemistry, and Biology 
\cite{rat,dif}.
Most prominently, directed transport by ratchet effects
has been scrutinized in a huge number of theoretical
works \cite{rat} due to its fascinating perspectives
with respect to basic Statistical Physics,
intracellular transport, and technological applications.
Also the experimental literature has substantially 
grown in recent years, demonstrating ratchet effects for
colloidal particles \cite{col},
Abrikosov vortices \cite{vor}
and Josephson vortices \cite{Maj03} in superconductors,
electrons in semiconductor heterostructures \cite{lin99},
Josephson phases across SQUIDs \cite{Ste05},
cold atoms \cite{men99},
granular gases \cite{mer04},
single cells \cite{cel},
and various magnetic systems \cite{mag}.
A related magnetic ratchet effect is also at
the basis of our present work:
The parallel magnetic stripes (domains)
of a ferrite garnet film (FGF)
are externally modulated
by an oscillating magnetic field,
resulting in a deterministic ratchet effect of
paramagnetic particles in the form
of directed transport perpendicularly
to the stripes.
The main result is the observation that
this ratchet effect triggers a giant enhancement of
the transversal (parallel to the stripes)
diffusion of the particles,
exceeding their field-free
diffusion
by almost four decades.
Moreover, by varying the
oscillation frequency, the transversal diffusion coefficient
exhibits a pronounced peak.
This anomalous behavior is explained
in terms of an analytically model.
In contrast to enhanced diffusion
in (effectively) one-dimensional models \cite{d1} 
and experiments \cite{d2}, 
this effect
is only possible in two dimensions and
is governed by a completely different physical
mechanism.
Also Galton-board type systems \cite{d3}
are very different from ours.

\begin{figure}
\epsfxsize=1.0\columnwidth
\epsfbox{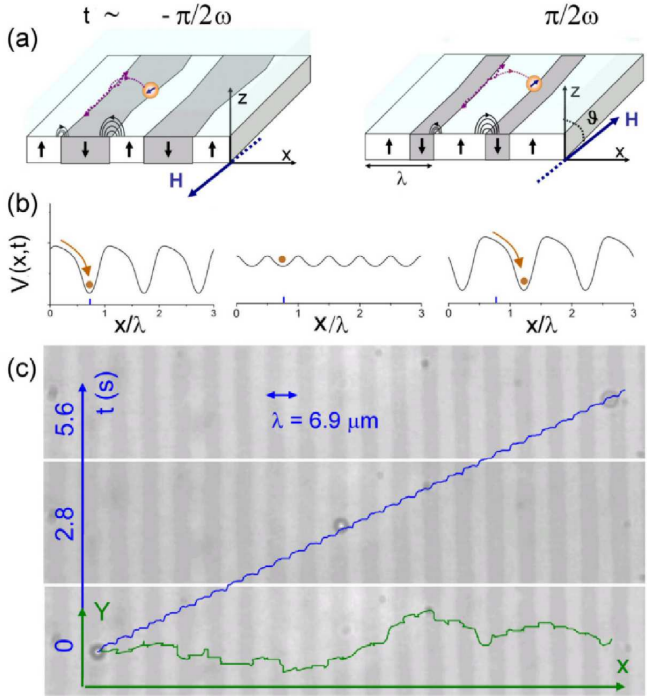} 
\caption{(Color online) (a) Schematic illustrations of a traveling paramagnetic
particle on top of the FGF, the stray field of the BWs, and
the oscillating external field $\bm{H}$ for two different times
$t = - \pi/2\omega$ and $t = \pi/2\omega$.
(b) Corresponding  potentials $V(x,t)$ (arb. units)
entailing particle motion in $x$-direction and calculated
from eq.(1) of Ref.~\cite{Mag}. The middle graph corresponds to $H=0$.
(c) Gray: three optical microscope snapshots of the FGF
and the traveling particle at $t=0$, $2.8$, and $5.6$ $s$
($\omega=18.8$ $s^{-1}$, $H=1500$ $A/m$, $\vartheta = \pi/7$).
Black (blue): corresponding particle trajectory $x(t)$.
Gray (green): corresponding path in the $x$-$y$-plane.
See also Video2 in~\cite{EPAPS}.}
\label{fig_1}
\end{figure}

In our experiments we use polystyrene 
paramagnetic particles (radius $a = 1.4$ $\mu m$,
volume susceptibility $\chi = 0.4$, Dynabeads M-270),
diluted in deionised water
(negligible particle-particle interaction),
and moving on top of an epitaxial grown
FGF with uniaxial  anisotropy~\cite{Tie00}.
As sketched in Fig. \ref{fig_1}(a), the FGF
exhibits periodic stripes of magnetic domains
with alternating up/down magnetization direction
(saturation magnetization $M_s = 1.7 \times 10^4$ $A/m$,
spatial periodicity $\lambda = 6.9$ $\mu m$).
The domains are separated by Bloch walls (BWs),
where the magnetic stray field $\bm{H}_{s}$
of the FGF is maximal.
Application of a spatially uniform
oscillating field with inclination
$\vartheta$ in the $x$-$z$ plane,
$\bm{H} \equiv H\, \sin(\omega t) (\sin \vartheta,0,\cos \vartheta)$,
has two main effects:
first, the $z$-component of $\bm{H}$
displaces the BWs by increasing
(resp. decreasing) the width of the domains with parallel
(resp. opposite) magnetization direction. 
The $x$-component of $\bm{H}$ breaks the symmetry of the 
potential as the BW array becomes an alternating sequence of strong and 
weak pinning sites according to where the local stray field is parallel
or antiparallel to the in-plane applied field,
respectively, see Fig. \ref{fig_1}(a).
Second, a paramagnetic particle acquires a magnetic moment
$\bm{m}=(4 a^3 \pi /3) \chi (\bm{H}+\bm{H}_{s})$,
entailing a magnetic force
$\bm{F} = \mu \nabla ((\bm{H}+\bm{H}_{s}) \cdot \bm{m} )) $,
where $\mu$ is the magnetic susceptibility of the medium.
As detailed in~\cite{Mag}, the particle
motion on top on the FGF is basically confined to
the $x$-$y$-plane and the net force
along the $x$-axis derives from a potential  $V(x,t)$ 
as indicated in Fig. \ref{fig_1}(b), while the
forces in $y$-direction are negligible
in a first approximation.
The spatially asymmetric ratchet-potential
$V(x,t)$ gives rise to a ratchet effect
\cite{rat},
``dragging'' along the particles in a deterministic
way (thermal noise is negligible) by
one spatial period $\lambda$ in $x$-direction
during one driving period $2\pi/\omega$.
We experimentally measure a maximum driving frequency of
$\omega_{max}\approx 200\, s^{-1}$, beyond which
the overdamped particles cannot follow the fast modulations.

Particles are tracked with a light microscope
($E400$, Nikon) equipped with a CCD camera (Basler)
working at 60 frames per second.
After recording $N$ independent trajectories $\{x_n(t),\, y_n(t)\}_{n=1}^N$,
we measure
the transversal diffusion coefficient in the
long-time limit,
$D_y=\lim_{t \to \infty}\langle [ y(t) - \langle y(t)\rangle ]^2 \rangle / 2t$,
where $\langle ... \rangle$ denotes an average over independent experiments.

\begin{figure}[b]
\epsfxsize=1.0\columnwidth
\epsfbox{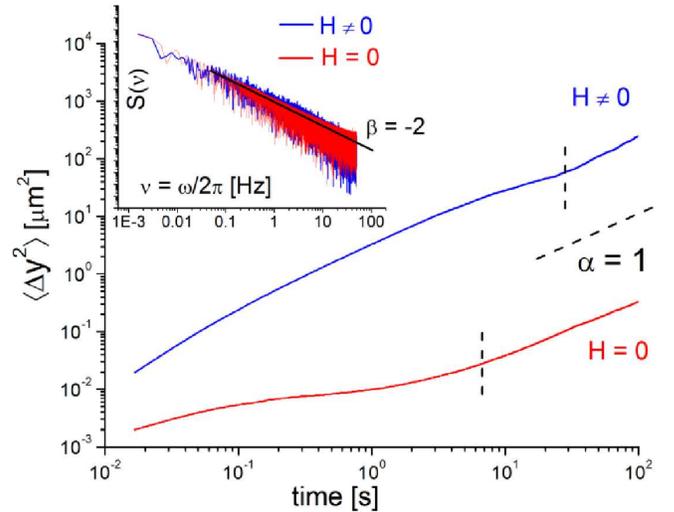} 
\caption{(Color online). Transversal mean square displacement
$\langle \Delta y^2 \rangle$ versus time (log-log-plot).
Black (blue): experimental results for the
same system as in Fig. \ref{fig_1}.
Gray (red): same but without external field.
The approach of a straight line with slope $\alpha=1$
for large times indicates convergence in $D_y$;
the curves starting from the dashed lines were used to calculate $D_y$.
Inset: The power spectra
$S(\nu):=|Y(\nu)|^2$, where $Y(\nu)$ is the Fourier transform
of one representative trajectory $y(t)$.
The convergence towards a straight line
with slope $\beta=-2$ for small $\nu$ indicates
normal (as opposed to anomalous) diffusion \cite{Gol06}.}
\label{fig_2}
\end{figure}

Without external field $\bm{H}$, the particles are pinned
by the BWs,
resulting in transversal thermal diffusion,
while longitudinal excursions from the BWs
are excluded (Video1 in~\cite{EPAPS}).
The quantitative details are provided with
Fig. \ref{fig_2}, yielding an unperturbed transversal diffusion
coefficient of $D_0 = 3.2 \times 10^{-4}$ $\mu m^2/s$,
which is almost three decades smaller than
the diffusion coefficient $D_{sf} = 0.09$ $\mu m^2/s$
measured on a stripe-free FGF~\cite{note1}.

\begin{figure}[t]
\epsfxsize=1.0\columnwidth
\epsfbox{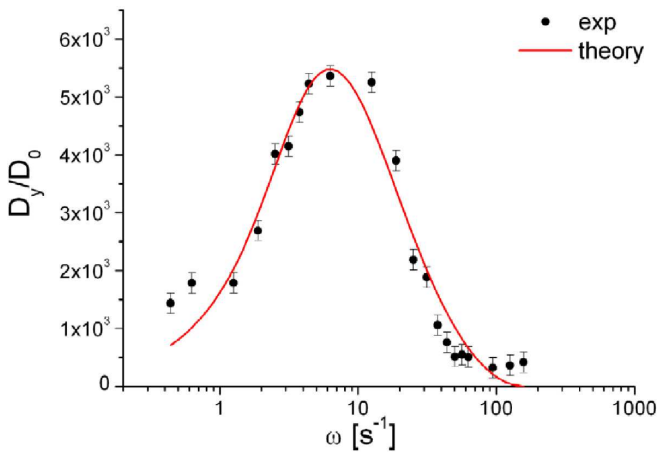} 
\caption{(Color online) Transversal diffusion coefficient $D_y$
versus driving frequency $\omega$ in units of
$D_0 = 3.2 \times 10^{-4}$ $\mu m^2/s$ (undriven diffusion).
Dots: Experimental findings for the same system as in Fig. \ref{fig_1}.
Line: Theoretical approximation (\ref{2}) with fit parameters
$\sigma=1.8 \pm 0.2$ $\mu m$ and $k=2.9 \pm 0.1$ $s^{-1}$.}
\label{fig_3}
\end{figure}

Upon application of the oscillating field $\bm{H}$,
the particles exhibit the above mentioned
ratchet effect in the $x$-direction.
The concomitant motion in $y$-direction
shows the typical qualitative features of
an unbiased random walk (Fig. \ref{fig_1}(c), Video2 in~\cite{EPAPS}).
Fig. \ref{fig_2} confirms quantitatively that we are
indeed dealing with a standard (normal) diffusion
process \cite{Gol06}, with a
mean square displacement
characterized by a power law behavior with exponent $\alpha=1$
for large $t$.
Striking enough, this diffusion
coefficient $D_y$ exhibits a pronounced peak as a
function of the driving frequency $\omega$,
exceeding the unperturbed diffusion $D_0$
by up to almost four orders of magnitude,
see Fig. \ref{fig_3}.
The rest of the paper is devoted to the
explanation of these findings.

\begin{figure}[b]
\epsfxsize=1.0\columnwidth
\epsfbox{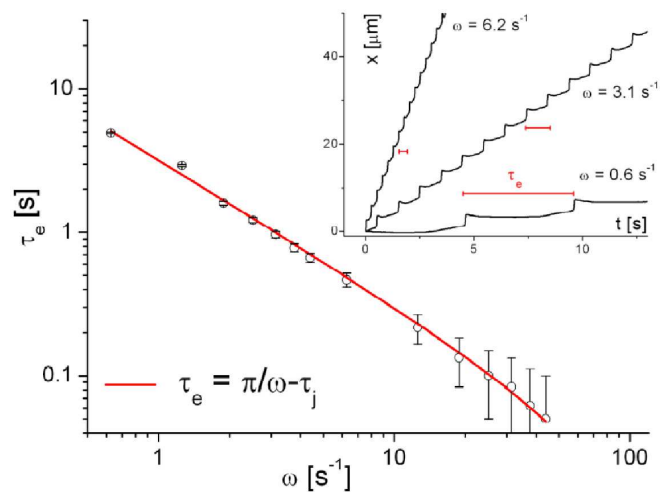} 
\caption{(Color online) Exploration time $\tau_e$ versus driving frequency $\omega$.
Circles: experimental results.
Line: Theoretical approximation $\tau_e = \pi/\omega-\tau_j$
with $\tau_j=0.02$ $s$. Inset: experimental particle trajectories
$x(t)$ for different driving frequencies $\omega$.
$\tau_e$ (red bars) indicates the exploration time 
between two jumps in $x$-direction, during which 
the particle moves along one given BW.}
\label{fig_4}
\end{figure}

According to Fig. \ref{fig_1}, a particle performs
a jump from one BW to the next
during every half-period $\tau =\pi/\omega$ of
the driving.
Furthermore, each jump is initiated by the
progressive disappearance of the potential 
well pinning the particle and is completed by
the subsequent relaxation towards a new potential minimum.
For driving frequencies $\omega$ well below the threshold 
$\omega_{max}$,
the duration $\tau_j$ of a jump is much
smaller than the half-period $\tau$
and turns out to be approximately independent of $\omega$.
In turn, for $\omega$ exceeding $\omega_{max}$, a
jump would take more than the total available half-period
$\tau$ and hence the entire ratchet mechanism
from Fig. \ref{fig_1} breaks down.
These considerations suggest the rough
estimate $\tau_j\approx \pi/\omega_{max}= 0.02\, s$,
in very good agreement with the experimental data
from Fig. \ref{fig_4}.
Altogether, within every half-period
$\tau$, the particle thus moves along one and
the same BW during an exploration time
of length $\tau_e=\tau-\tau_j$, and then
jumps to the next BW.

Within the diffraction limit ($\approx 200$ $nm$)
of our optics, the BWs appear as practically
straight borderlines in Fig \ref{fig_1}(c).
However, it is known that on smaller scales 
they do exhibit notable random undulations,
caused by pinning sites and other inhomogeneities of the FGF.
As shown in Ref.~\cite{undu}, the amplitude and 
bend period of the undulations are not reproducible 
upon reapplication of the field.
As a consequence, the random undulations of 
different BWs are independent of each other.
Due to similar reasons, also the magnetic
stray field, experienced by a particle when moving 
along a BW, exhibits random variations.

The simplest quantitative modeling
consists in assuming an overdamped relaxation dynamics
$\eta\dot y(t)=-U'(y(t))$, where $U(y)$ is
a random magnetic potential, negligibly weak
with regard to jumps between the BWs
($x$-direction),
but not with regard to the dynamics in
$y$-direction (along the BWs).
Thermal noise is neglected in view of our
experimental finding that the field-free
diffusion $D_0$ along a BW is much smaller 
than thermal diffusion $D_{sf}$ on a stripe-free FGF.
The damping coefficient $\eta$
can be estimated via Einstein's formula
$D_{sf}=k_BT/\eta$, with $k_B$ Boltzmann's constant
and $T\approx 300\,$K,
yielding $\eta\approx 0.05\, $mg/s.
Jumps between BWs are modeled by resetting/resampling the
random potential $U(y)$, while keeping $y(t)$ fixed during the 
corresponding jump time $\tau_j$.
To be specific, we focus on the simplest possible
example, namely a potential $U(y)$ composed of
parabolic pieces, each piece (index $i$)
exhibiting the same curvature $\kappa>0$,
but with randomly sampled extensions 
to the right and to the left of its minimum $y_i$.
Thus a seed $y(0)$ within the $i$-th
parabolic piece evolves according to
$\dot y(t)= - k\,(y(t)-y_i)$ with $k:=\kappa/\eta$.
For the displacement after half a period $\tau$
one thus obtains $y(\tau)-y(0)= \xi (1-e^{-k\tau_e})$
with $\xi:= y_i-y(0)$.
Similarly, one obtains for the displacement after 
$P$ half-periods $y(P\tau)-y(0)=(1-e^{-k\tau_e}) \sum_{p=1}^P \xi_p$,
where the resampling of the random potential $U(y)$ 
after each jump
implies that the $\xi_p$ are independent, identically
distributed random numbers of zero average and
a finite variance $\sigma^2$.
This leads to:
\begin{equation}
D_y=\frac{\sigma^2}{2 \tau}\, \left[1-e^{-k (\tau-\tau_j)}\right]^2 \ ,
\label{2}
\end{equation}
where $\sigma$ and $k$ are fit parameters.
The former can be readily identified with the
characteristic length scale of the BW undulations,
while $k^{-1}$ gives the relaxation time scale.
Fig. \ref{fig_3} demonstrates very good
agreement of (\ref{2})
with the experimental observations,
with $\sigma$ being reasonably of the order of the domain width
($\lambda$/2)
and $k^{-1}$ matching the time scale of the applied forcing 
when a significant enhancement of $D_y$ occurs.
Admitting more than a single curvature of the
parabolic pieces yields an analytical result containing
(\ref{2}) as special case, and thus fitting the
experiment even better.
Further analytical solutions are possible
for piecewise sinusoidal instead of
piecewise parabolic random potentials $U(y)$,
fitting the experiment practically equally well.
The details of $U(y)$ thus seem to matter very little.

For small $\omega$, the asymptotics $D_y\sim\omega$
of (\ref{2}) reflects the fact that the particles have ample
time to relax into the next potential minimum along
every BW, but then have to await the next jump.
Conversely, for $\omega \to \omega_{max}$, the particles have
almost no time to
explore the BWs and thus $D_y\sim (\omega_{max}-\omega)^2$
according to (\ref{2}).
It follows that there must be some intermediate $\omega$
at which $D_y$ exhibits a maximum.
The vanishing of (\ref{2}) at $\omega=\omega_{max}$ is 
non-realistic, and is due to
neglecting any motion 
in $y$-direction during the jump times $\tau_j$.
Within our model, the field-free diffusion $D_0$ corresponds
to a single, fixed random potential $U(y)$, and the
stripe-free diffusion $D_{sf}$ to a constant $U(y)$.

Experimentally, variations of the driving amplitude $H$
were restricted to the regime $0.06\, M_s < H < 0.15\, M_s$.
Within this regime, $D_y$ did not notably depend on $H$,
in agreement with the theory.
For smaller $H$, the particles were not able to
jump, while larger $H$ lead to
irreversible deformations of the stripes
and inhibition of particle hopping between BWs.
We also explored the effect of high bending stripe
deformation (i.e. ``zig-zag'' stripe patterns),
which leads to a counterintuitive 
decrease of the transversal diffusion
coefficient $D_y$ since the particles funnel
into narrow trajectories 
(see Fig. 2 in \cite{EPAPS}).
The natural explanation within our model
follows by superimposing to the random potentials 
$U(y)$ one and the same periodic potential for all BWs,
taking into account that they all exhibit the same 
``zig-zag'' pattern.

To conclude, we observed giant transversal
diffusion of paramagnetic colloidal particles
on a garnet film, triggered by a longitudinal
external driving and a concomitant longitudinal 
ratchet effect.
We interpret this as the signature of a
field induced undulation instability in the
magnetic domain structures,
invoking the presence of a time-refreshed 
disordered potential landscape in the 
magnetic pattern.
The understanding and control of diffusion
along the lines of our present work is 
a key issue in many technological and 
biophysical contexts \cite{d1,d2,d3}.

We acknowledge discussions with J. M. Sancho and T. M. Fischer.
P.T. was supported by the program ``Juan de la Cierva'' (JCI-2009�04192).
P.T. and F. S. 
acknowledge financial support
by MEC (FIS2006-03525) and DURSI
(2009SGR1055);
P.R. by Deutsche Forschungsgemeinschaft (SFB 613),
and T.H.J. by The Research Council of Norway.


\begin{thebibliography}{99}
\bibitem{note1}
The same FGF under an external field $H>M_s$ such that the surface
has no magnetic stripes.
\end{thebibliography}
\end{document}